# Building Near-Real-Time Processing Pipelines with the Spark-MPI Platform

Nikolay Malitsky
*NSLS-II Department*
*Brookhaven National Laboratory*
Upton, NY 11973, USA
malitsky@bnl.gov

Aashish Chaudhary
*Scientific Computing*
*Kitware, Inc.*
Clifton Park, NY 12065, USA
aashish.chaudhary@kitware.com

Sebastien Jourdain
*Scientific Computing*
*Kitware, Inc.*
Santa Fe, NM 87501, USA
sebastien.jourdain@kitware.com

Matt Cowan
*NSLS-II Department*
*Brookhaven National Laboratory*
Upton, NY 11973, USA
cowan@bnl.gov

Patrick O'Leary
*Scientific Computing*
*Kitware, Inc.*
Santa Fe, NM 87501, USA
patrick.oleary@kitware.com

Marcus Hanwell
*Scientific Computing*
*Kitware, Inc.*
Clifton Park, NY 12065, USA
marcus.hanwell@kitware.com

Kerstin Kleese Van Dam
*Computational Science Initiative*
*Brookhaven National Laboratory*
Upton, NY 11973, USA
kleese@bnl.gov

*Abstract*—Advances in detectors and computational technologies provide new opportunities for applied research and the fundamental sciences. Concurrently, dramatic increases in the three V's (Volume, Velocity, and Variety) of experimental data and the scale of computational tasks produced the demand for new real-time processing systems at experimental facilities. Recently, this demand was addressed by the Spark-MPI approach connecting the Spark data-intensive platform with the MPI high-performance framework. In contrast with existing data management and analytics systems, Spark introduced a new middleware based on resilient distributed datasets (RDDs), which decoupled various data sources from high-level processing algorithms. The RDD middleware significantly advanced the scope of data-intensive applications, spreading from SQL queries to machine learning to graph processing. Spark-MPI further extended the Spark ecosystem with the MPI applications using the Process Management Interface. The paper explores this integrated platform within the context of online ptychographic and tomographic reconstruction pipelines.

*Keywords—streaming, high-performance, data analysis, experimental facility, Spark, MPI*

## I. INTRODUCTION

The fourth paradigm of data-intensive science [1], coined by Jim Gray, rapidly became a major driving concept of multiple application domains ranging from nanotechnology to cosmology, encompassing and generating large-scale devices such as light sources, neutron sources, fusion reactors, and cutting edge telescopes [2-4]. Data-intensive applications however required the reconsideration of many existing technical approaches and led to the development of a new Big Data ecosystem without intersections with the HPC computer-intensive environment built within the previous computational paradigm. As a result, the existing impedance mismatch between data-intensive and compute-intensive ecosystems introduced the demand for integrated platforms addressing synergistic challenges of large-scale data-driven scientific applications [5].

The initial landscape of Big Data technologies was designed after Google's I/O stack, including the MapReduce processing framework [6]. Addressing immediate data-intensive applications, the MapReduce computing platform played a significant role in the expansion of the Apache ecosystem and the formation of a reference level for developing new technologies. This level has been further elevated by the Spark programming model [7] based on resilient distributed datasets (RDDs) and a comprehensive collection of RDD transformations and actions. The new model of the Spark computing platform significantly advanced and extended the scope of data-intensive applications, spreading from SQL queries to machine learning to graph processing. Moreover, Spark facilitated the implementation of the lambda architecture [8] by adding the Spark Streaming module [9].

The combination of a data-intensive processing framework with a consolidated collection of diverse data analysis algorithms offered by Spark represents a strong asset for its application in large-scale scientific experiments and computational projects. The inherited MapReduce embarrassingly parallel model however introduces conceptual obstacles for integrating the HPC scientific algorithms that were originally developed within the Message Passing Interface (MPI) model [10]. In addition, the complexity of advanced data-intensive applications steadily elevated them towards a compute-intensive domain generating a new heterogeneous collection of distributed processing frameworks, such as GraphLab [11], DistBelief [12], TensorFlow [13], and Gorila [14]. As with any standard evolutionary spiral, a variety and growing number of different approaches eventually raised the question of their consolidation.

The integration of data-intensive and compute-intensive ecosystems has been addressed by several projects. For example, Geoffrey Fox and colleagues provided one of the most comprehensive overviews of the Big Data and HPC domains. Their application analysis [15] was based on several surveys, such as the NIST Big Data Public Working Group and

This research was funded by the DOE ASCR SBIR program under Contract No. DE-SC0017133

NRC reports, including multiple application domains: energy, astronomy and physics, climate, and others. Another paper by this team [16] explored the Apache Big Data Stack (ABDS) and HPC technologies using a list of 289 software projects. Their analysis explicitly highlighted common and complementary approaches of two ecosystems through 21 layers of a reference HPC-ABDS architecture.

This paper addresses the existing mismatch between Big Data and HPC applications by proposing the incremental consolidation of Big Data processing frameworks on HPC infrastructure within the context of emerging tasks. As the first step on this path, the paper presents the Spark-MPI integrated platform connecting the Spark and MPI technologies for building data-intensive high-performance data processing pipelines for experimental facilities. The paper further develops and assesses this direction within the context of ptychographic and tomographic applications.

## II. SPARK-MPI

The Spark-MPI approach was derived from the CaffeOnSpark [17] and Sharp-Spark [18] projects that were developed within the context of two applications: deep learning and ptychographic reconstruction. CaffeOnSpark extended the Spark driver-worker model with an inter-worker interface based on RDMA over InfiniBand. The Sharp-Spark approach followed the CaffeOnSpark peer-to-peer model and augmented it with an RDMA address exchange server that significantly facilitated the initialization phase responsible for establishing inter-worker connections. Within the MPI ecosystem, the same task is solved using the Process Management Interface (PMI[19]). As a result, the Sharp-Spark project provided a natural transition to the Spark-MPI approach based on the PMI server. Fig. 1 shows the conceptual diagram of this approach.

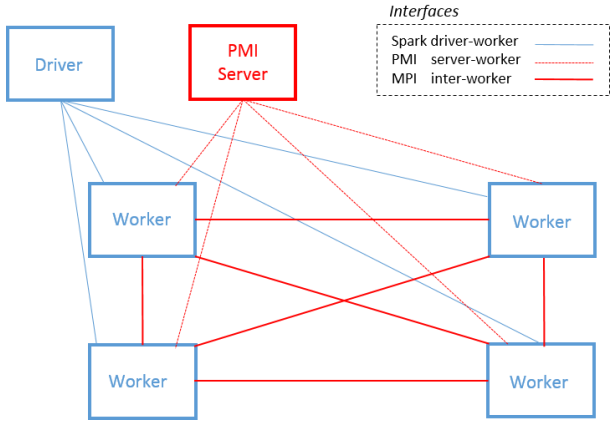

Fig. 1. The Spark-MPI approach

Spark-MPI encompasses three interfaces. Specifically, it complements the Spark conventional driver-worker model with the PMI server-worker interface for establishing MPI inter-worker communications. The Spark driver is a main program that creates RDDs, applies transformations and actions, and submits the corresponding requests to a cluster manager that launches executors on distributed workers for processing the RDD operations. Spark is a part of the Apache Hadoop suite of Big Data technologies developed for a shared-nothing cluster environment. Therefore, Spark applications can run with the Hadoop YARN cluster manager [20] and Hadoop filesystem (HDFS). The scope of the Spark runtime environment however is not limited by the Hadoop ecosystem and can be extended with other cluster managers, such as Apache Mesos [21] or a Standalone process manager. Furthermore, the same applications can be deployed on an HPC cluster platform managed by Simple Linux Utility for Resource Management (SLURM [22]) by embedding the Spark Standalone cluster manager as shown in the sbatch script in Fig. 2. After submitting this script, SLURM allocates workers, starts the Spark Standalone master and slaves, and launches the Spark application.

```
start-master.sh
srun spark-class org.apache.spark.deploy.worker.Worker spark://$SLURMD_NODENAME:7077 &
spark-submit  --master spark://$SLURMD_NODENAME:7077 ./collect.py 10
```

Fig 2. A SLURM batch script for running a Spark application

In the MPI frameworks, the implementation of the PMI interface consists of two parts: client and server. The client side (e.g., Simple PMI) is linked with the MPI library. And the server side is associated with a corresponding process manager (e.g., Hydra or SLURM). As a result, PMI enables reusing the process manager for exchanging information needed to connect processes together. This interface is defined around the concept of key-value space (KVS), containing a set of (key, value) pairs. Processes acquire access to one or more KVS's through the PMI library and can perform put/get operations on them. Synchronization is provided in a scalable way via the barrier operation that assures that the necessary puts have been done before attempting the corresponding gets. The present PMI version was released as part of the MPICH project and accepted by other MPI implementations and external process managers.

Spark-MPI approach was validated with the primary internal process manager, Hydra, which is used by two MPI projects, MPICH and MVAPICH. As shown in Fig. 3, Hydra provides two major functions: process launching and PMI-based information exchange.

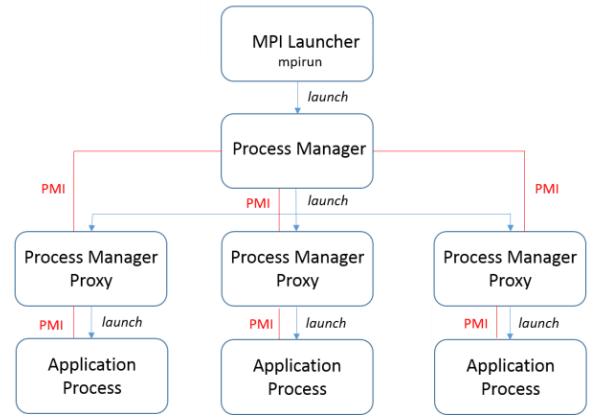

Fig. 3. Hydra process manager

The Process Manager (PM) is started by the MPI executable mpirun on the launch node and subsequently spawns multiple PM proxies on each node of the application environment. Following the Lernaean Hydra story, each proxy can be setup in a hierarchical fashion and acts as a process manager for its sub-tree. Proxies launch application processes and implement the PMI interface for exchanging information between application processes and the main process manager. The implementation of the Hydra-based PMI server required only a few changes in order to suppress the launching of application processes. For launching the PMI server, the SLURM batch script in Fig. 2 needs to be extended with a few commands as shown in Fig. 4.

```
scontrol show hostname $SLURM_JOB_NODELIST | paste -d'\n' -s > hosts
pmiserv -f hosts &

start-master.sh
srun spark-class org.apache.spark.deploy.worker.Worker spark://$SLURMD_NODENAME:7077 &
spark-submit --master spark://$SLURMD_NODENAME:7077 ./allreduce.py 10
```

Fig. 4. The SLURM batch script for running the Spark-MPI application

Following the Sharp-Spark project, the Spark-MPI approach can be illustrated by the AllReduce benchmark implemented with two alternative scripts, collect.py (see Fig. 5) and allreduce.py (see Fig. 6).

```
sc = SparkContext()

partitions = int(sys.argv[1]) if len(sys.argv) > 1 else 1
n = 2*1000000

env = []
for id in range(0, partitions):
    kvs = { 'n' : n, 'rank' : id}
    env.append(kvs)

# Create the RDD instance
rdd = sc.parallelize(env, partitions)

def sendbuf(kvs):
    rank = kvs['rank']
    n = kvs['n']

    sendbuf = np.arange(n, dtype=np.float32)
    sendbuf[n-1] = 5.0;

    t1 = datetime.now()
    out = { 'rank' : rank, 't1' : t1, 'buffer' : sendbuf }
    return out

# Collect buffers from distributed workers
results = rdd.map(sendbuf).collect()

# Print timing
t2 = datetime.now()
for out in results:
    print ("rank: ", out['rank'], ", buffer: ", out['buffer'][n-1], ", processing time: ", t2 - out['t1'])
```

Figure 5. The collect.py script based on the Spark driver-worker model

The benchmark compares the performance of these two approaches for summing large arrays of floats across the Spark workers. Both scripts use the same RDD instance that is created in the beginning of this script. In Figure 5, the script collects arrays from different workers using the Spark driver-worker interface and sums them on the Spark driver. The allreduce.py script is based on the Spark-MPI approach and uses the MPI Allreduce method provided by the Python module mpi4py. As shown in this script, it requires defining only two environmental variables, PMI_PORT and PMI_ID, for connecting the Spark worker with the PMI server. The corresponding results for different number of nodes are compiled in Table I.

The benchmark results highlight the substantial performance speedup of the proposed Spark-MPI approach for compute-intensive applications. The last entry of Table I includes results produced by the MVAPICH application running with the ch3:sock device. It shows the limits of the existing protocols, such as gRPC/Ethernet, used in distributed deep learning frameworks and identifies it as an area for future upgrades with the MPI framework.

```
sc = SparkContext()

partitions = int(sys.argv[1]) if len(sys.argv) > 1 else 1

# Create the rdd collection associated with the MPI workers
env = [id for id in range(partitions)]
rdd = sc.parallelize(env, partitions)

# Define the MPI application
def allreduce(kvs):

    # unsetenv the SLURM env. variables
    del os.environ["PMI_FD"]

    # define env. variables of the Spark-PMI server
    hostname = socket.gethostname()
    hydra_proxy_port = os.getenv("HYDRA_PROXY_PORT")
    pmi_port = hostname + ":" + hydra_proxy_port

    os.environ["PMI_PORT"] = pmi_port
    os.environ["PMI_ID"]   = os.getenv("SLURM_NODEID")

    # import and initialize MPI
    from mpi4py import MPI
    comm = MPI.COMM_WORLD
    rank = comm.Get_rank()

    # image
    n = 2*1000000
    sendbuf = np.arange(n, dtype=np.float32)
    recvbuf = np.arange(n, dtype=np.float32)

    t1 = datetime.now()
    for i in range(10):

        sendbuf[n-1] = i;
        comm.Allreduce(sendbuf, recvbuf, op=MPI.SUM)

    t2 = datetime.now()
    out = { 'rank' : rank, 'time' : (t2-t1)/10, 'sum' : recvbuf[n-1] }
    return out

# Run MPI application on Spark workers and collect the results
results = rdd.map(allreduce).collect()

for out in results:
    print ("rank: ", out['rank'], ", sum: ", out['sum'], ", processing time: ", out['time'])
```

Fig. 6. The allreduce.py script based on the Spark-MPI approach

TABLE I. PERFORMANCE COMPARISON: SPARK (OVER ETHERNET), SPARK-MPI OVER INFINIBAND, AND MPI OVER ETHERNET

| Approach | Time (s) vs Number of Nodes | | | |
|---|---|---|---|---|
| | 2 | 4 | 8 | 10 |
| Spark/Ethernet driver-worker model (see Fig. 5) | 0.20 | 0.37 | 0.95 | 1.12 |
| Spark-MPI approach based on MVAPICH2/InfiniBad (see Fig. 6) | 0.0036 | 0.0049 | 0.0060 | 0.0097 |
| MVAPICH /Ethernet | 0.07 | 0.14 | 0.31 | 0.36 |

The integration of the MPI framework with the Spark RDD middleware immediately provides a connection between MPI applications and different types of distributed data sources including major databases and file systems. Furthermore, the Spark Streaming module [9] reused and extended the RDD-based batch processing framework with a new programming abstraction called discretized stream, a sequence of RDDs, processed by micro-batch jobs. These new batches are created at regular time intervals. Similar to batch applications, streams can be ingested from multiple data sources like Kafka, Flume, Kinesis and TCP sockets.

In this paper, the Spark-MPI approach was evaluated with the Kafka streaming platform, an Apache open source project that was originally developed at LinkedIn [23]. Kafka is a scalable message broker providing topics with streams of records that can be produced and consumed by multiple clients. A topic is divided into partitions adding horizontal scalability and parallelism. Each partition is physically stored as a series of segment files. A message in Kafka consists of a key and a value, which are untyped variable-length byte strings. Messages within a partition are completely ordered. Kafka however does not provide ordering guarantee across different partitions.

In Spark Streaming, different input sources can be divided into basic and advanced categories. The connection with the basic sources, such as file systems or sockets, is directly provided from the StreamingContext. Kafka belongs to the second category of sources that are supported via extra utility classes, such as KafkaUtils, implemented in several programming languages including Java and Python. KafkaUtils provides two options for receiving messages from Kafka topics. First, it can create an input stream that is automatically updated by the Spark Streaming framework. As an intermediate step, this paper considered the second more flexible option that explicitly creates an RDD using offset ranges. The corresponding scenario is described by a conceptual diagram (see Fig. 7) and a Python method (see Fig. 8).

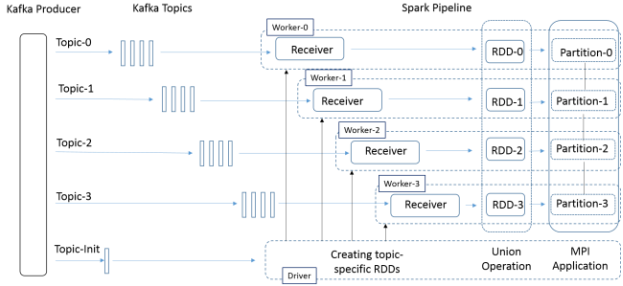

Fig 7. Streaming demo with the Spark-MPI approach

```
def run_batch(sc, partitions, start, until):

    # create the Kafka topic-specific RDDs

    kafkaParams = {"metadata.broker.list": "localhost:9092"}

    kafkaRDDs = []
    for j in range(partitions):
        topic = "topic-" + str(j)
        offset = OffsetRange(topic, 0, start, until)
        offsets = [offset]
        topicRDD = KafkaUtils.createRDD(sc, kafkaParams, offsets,
                                        valueDecoder=valueDecoder)
        kafkaRDDs.append(topicRDD)

    # combine topic-specific RDDs
    unionRDD = sc.union(kafkaRDDs)
    numParts = unionRDD.getNumPartitions()
    print("number of partitions: ", numParts)

    # run the MPI application
    unionRDD.mapPartitions(allreduce).count()
```

Figure 8. The run_batch method

According to this scenario, the input stream represents a sequence of micro-batches. The Spark driver waits for a topic-init record and processes each micro-batch with the run_batch method (see Fig. 8). At the beginning of this method, the Kafka data are ingested into the Spark platform as the Kafka RDDs. To achieve a higher level of parallelism, records of micro-batches are divided into topics that are consumed by Kafka Receivers on distributed Spark workers. Each Kafka Receiver creates a topic-specific RDD and the Spark driver logically combines them together with a union operation. As a result, it prepares a distributed RDD to be processed with the MPI application as shown in the previous allreduce.py script (see Fig. 6).

### III. PTYCHOGRAPHIC APPLICATION

Ptychography is one of the essential image reconstruction techniques used in light source facilities. It was originally proposed for electron microscopy [24] and lately applied to X-ray imaging [25-26]. The method consists of measuring multiple diffraction patterns by scanning a finite illumination (also called the probe) on an extended specimen (the object). The redundant information encoded in overlapping illuminated regions is then used for reconstructing the sample transmission function. Specifically, under the Born and paraxial approximations, the measured diffraction pattern for the $j$th scan position can be expressed as:

$$I_j(\mathbf{q}) = |F\psi_j|^2, \qquad (1)$$

where $\mathbf{F}$ denotes Fourier transformation, q is a reciprocal space coordinate, and $\psi_j$ represents the wave at the exit of the object O illuminated by the probe P:

$$\psi_j = P(\mathbf{r}-\mathbf{r}_j)O(\mathbf{r}) \qquad (2)$$

Then, the object and probe functions can be computed from the minimization of the distance $\|\Psi - \Psi^0\|^2$ as [27]:

$$\varepsilon = \|\Psi - \Psi^0\|^2 = \sum_j \sum_r |\psi_j(\mathbf{r}) - P^0(\mathbf{r}-\mathbf{r}_j)O^0(\mathbf{r})|^2 \qquad (3)$$

$$\frac{\partial \varepsilon}{\partial P^0} = 0: \ P^0(\mathbf{r}) = \frac{\sum_j \psi_j(\mathbf{r}+\mathbf{r}_j)O^*(\mathbf{r}+\mathbf{r}_j)}{\sum_j |O(\mathbf{r}+\mathbf{r}_j)|^2} \qquad (4)$$

$$\frac{\partial \varepsilon}{\partial O^0} = 0: \ O^0(\mathbf{r}) = \frac{\sum_j \psi_j(\mathbf{r})P^*(\mathbf{r}-\mathbf{r}_j)}{\sum_j |P(\mathbf{r}-\mathbf{r}_j)|^2} \qquad (5)$$

These minimization conditions need to be augmented with the modulus constraint (1) and included in the iteration loop. For example, the comprehensive overview of different iterative algorithms is provided by Klaus Giewekemeyr [28]. At this time, the difference map [29] is considered as one of the most generic and efficient approach to address these types of imaging problems. It finds a solution in the intersection of two constraint sets using the difference of corresponding projection operators, $\pi_1$ and $\pi_2$, composed with associated maps, $f_1$ and $f_2$:

$$\psi^{n+1} = \psi^n + \beta \ \Delta(\psi^n)$$
$$\Delta = \pi_1 \circ f_2 - \pi_2 \circ f_1 \qquad (6)$$
$$f_i(\psi) = (1 + \gamma_i)\pi_i(\psi) - \gamma_i\psi$$

where $\gamma_{1,2}$ are relaxation parameters. In the context of the ptychographic applications, these projection operators are associated with the modulus (1) and overlap (2) constraints. By selecting different values of relaxation parameters, the difference map (6) can be specialized to different variants of phase retrieval methods and hybrid projection-reflection (HPR) algorithms. Further developing HPR, Russel Luke [30] introduced the relaxed averaged alternating reflections (RAAR) approach:

$$\psi^{n+1} = [2\beta \pi_2\pi_1 + (1 - 2\beta) \pi_1 + \beta(1 - \pi_2)]\psi^n \quad (7)$$

The RAAR algorithm was implemented in the SHARP program [31] at the Center for Advanced Mathematics for Research Applications (CAMERA).

SHARP is a high-performance distributed ptychographic solver using GPU kernels and the MPI protocol. Since most equations with the exception of (4) and (5) are framewise intrinsically independent, the ptychographic application is naturally parallelized by dividing a set of data frames among multiple GPUs. Then, for updating a probe and an object, the partial summations of (4) and (5) are combined across distributed nodes with the MPI Allreduce method as shown in Fig. 9

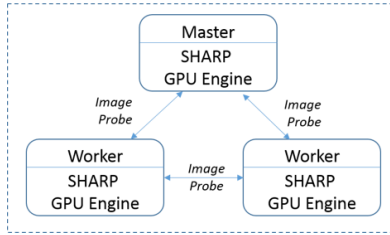

Fig. 9. MPI communication model of the SHARP solver

The SHARP multi-GPU approach significantly boosted the performance of ptychographic applications and immediately highlighted the direction associated with the development of near-real-time processing pipelines. The direct deployment of the SHARP program on the Spark Streaming platform however was prevented by Spark computational model constrains. To overcome this issue, the Sharp-Spark application [18] replaced the SHARP MPI-based communicator with a specialized version based on the Spark inter-worker extension developed by the Yahoo Big ML team within the CaffeOnSpark project [17]. In contrast with the Sharp-Spark application, the Spark-MPI approach provides a generic solution that does not require any changes in the original MPI programs. Furthermore, the corresponding composite data-intensive high-performance applications can be deployed on various systems using different communication protocols.

For benchmarking the Spark-MPI approach, we used the same simulation-based application from the Sharp-Spark project [18]. Fig. 10 shows object phases reconstructed from 512 detector frames after 100 iterations. In current experiments, the interval between scan points takes approximately 50 ms, in other words 25 seconds for 512 frames. As shown in Table II, the Spark-MPI application demonstrated the feasibility of near-real-time scenario. This direction is especially important from the perspective of a new category of emerging four-dimensional applications. One of them is a tomographic experiment that combines series of ptychographic projections generated at different angles of object rotation. In this experiment, each ptychographic projection is reconstructed from tens of thousands of detector frames and the MPI multi-GPU version becomes critical for addressing the GPU memory challenges of these applications.

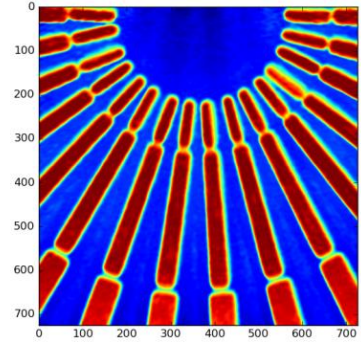

Fig. 10. Object phases

TABLE II. BENCHMARK RESULTS OF THE SHARP-NSLS2 APPLICATION PRODUCED ON A CLUSTER (GPU: TESLA K80)

| Application | Time (s) vs Number of Nodes | | |
| --- | --- | --- | --- |
| | 1 | 2 | 4 |
| SHARP-NSLS2 [18] | 22.7s | 13.6 | 8.6 |

IV. TOMOGRAPHIC APPLICATION

Tomography is one of the imaging techniques used in radiology, materials science, geophysics and many other domains of science. In tomography, we capture a series of images through the utilization of any penetrative wave. In the field of material science, tomography enables three-dimensional (3D) characterization of materials at the nano- and mesoscale. For our work, we are using the publicly accessible data referred to in the paper of Levin and colleagues [32]. The formatted data are in 16-bit tiff image format with sizes ranging from a few megabytes (MB) to a few gigabytes (GB).

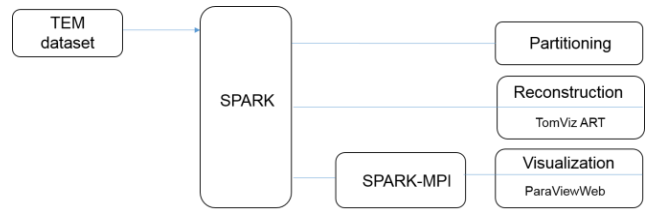

Fig. 11. Spark-MPI pipeline for reconstruction of TEM data

The conceptual workflow for our tomography reconstruction example is as follows (see Fig. 11):

1. Load a transmission electron microscopy (TEM) dataset into RDD format,
2. Partition the RDD dataset appropriately for performing a reconstruction algorithm,

3. Perform reconstruction on each partition of dataset in parallel, and
4. Gather the resulting reconstructed 3D dataset and render it using ParaView and ParaViewWeb

```
def processPartition(idx, iterator):
    // Reconstruction helper
    def parallelRay(Nside, pixelWidth, angles, Nray, rayWidth):
        // Call Recon_ART.py

    // Reconstruction
    tiltSeries = scalars_array3d
    tiltAngles = range(-sizeZ + 1, sizeZ, 2)
    (Nslice, Nray, Nproj) = tiltSeries.shape
    Niter = 1
    A = parallelRay(Nray, 1.0, tiltAngles, Nray, 1.0) // A is a sparse matrix
    recon = np.empty([Nslice, Nray, Nray], dtype=float, order='F')
    A = A.todense()
    (Nrow, Ncol) = A.shape
    rowInnerProduct = np.zeros(Nrow)
    row = np.zeros(Ncol)
    f = np.zeros(Ncol) // Placeholder for 2d image
    beta = 1.0

// Calculate row inner product
    for j in range(Nrow):
        row[:] = A[j, ].copy()
        rowInnerProduct[j] = np.dot(row, row)
    for s in range(Nslice):
        f[:] = 0
        b = tiltSeries[s, :, :].transpose().flatten()
        for i in range(Niter):
            for j in range(Nrow):
                row[:] = A[j, ].copy()
                row_f_product = np.dot(row, f)
                a = (b[j] - row_f_product) / rowInnerProduct[j]
                f = f + row * a * beta
        recon[s, :, :] = f.reshape((Nray, Nray))
    parallelVisualize(recon)
```

Fig. 12. ART reconstruction approach

In the Spark-MPI integration, we launch a Hydra-based custom server with the desired number of processes and then initiate MPI ranks in the Spark context using Hydra environment variables.

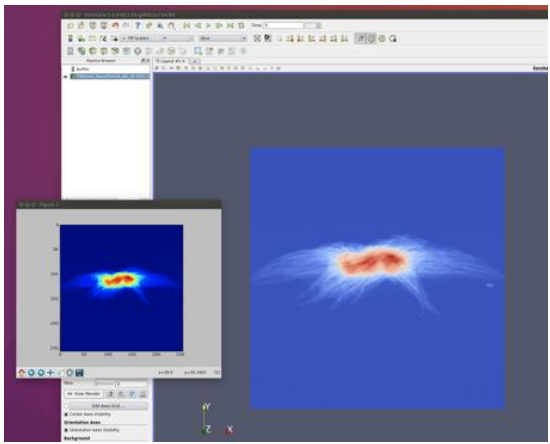

Figure 13. Verification slice.

For the first step, we explored several options to perform extract-transform-load (ETL) on the TEM dataset. First, we leveraged the fastparquet library to perform TIFF to Parquet format conversion and then used the Parquet format for creating an RDD format dataset. Figure 13 shows a matplotlib plot that uses the RDD format dataset to render a slice for visual verification. The other options we pursued utilized NumPy with the Spark parallelized collections utility. In the NumPy-based workflow, the data are read as NumPy array and distributed using Spark Context's parallelize method on an existing collection. The NumPy array is copied to form a distributed dataset that supports parallel operations.

Once the data is in Spark context as RDD, we repartition the data to ensure the neighboring pixel are in the same partition for reconstruction algorithm.

For the third step of our TEM example, we chose the algebraic reconstruction technique (ART) implemented in TomViz [33] and executed it within the processPartition method on each partition of data in parallel using Spark map-collect paradigm (see Fig. 12 and Fig. 14).

Finally, we perform the rendering and visualization utilizing the ParaView server that employs the MPI environment for parallelization and the ParaViewWeb client Visualizer enabling interactive viewing in a Browser.

```
def parallelVisualize(recon):
    (iSize, jSize, kSize) = recon.shape
    os.environ["PMI_PORT"] = pmi_port
    os.environ["PMI_ID"] = str(idx)
    os.environ["PV_ALLOW_BATCH_INTERACTION"] = "1"
    os.environ["DISPLAY"] = ":0"

    // Convert reconstruction array into VTK format
    arr = recon.ravel(order='A')
    vtkarray = numpy_support.numpy_to_vtk(arr)
    vtkarray.SetName('Scalars')

    dataset = vtkImageData()
    minX = 0
    maxX = 0
    for i in range(idx + 1):
        minX = maxX
        maxX += xSizes[i]
    dataset.SetExtent(minX, maxX - 1, 0, sizeY - 1, 0, sizeY - 1)
    dataset.GetPointData().SetScalars(vtkarray)
    vtkDistributedTrivialProducer.SetGlobalOutput('Spark', dataset)

    // Import VTK and ParaView modules here
    pm = vtkProcessModule.GetProcessModule()
    class _VisualizerServer(pv_wamp.PVServerProtocol):
        dataDir = '/data'
        groupRegex = "[0-9]+\\.[0-9]+\\.|[0-9]+\\."
        excludeRegex = "^\\.|~$|\\$"
        allReaders = True
        viewportScale=1.0
        viewportMaxWidth=2560
        viewportMaxHeight=1440
        def initialize(self):
            .....
            args = Options()
            if pm.GetPartitionId() == 0:
                producer = simple.DistributedTrivialProducer()
                producer.UpdateDataset = ''
                producer.UpdateDataset = 'Spark'
                producer.WholeExtent = [0, sizeX - 1, 0, sizeY - 1, 0, sizeY - 1]
                server.start_webserver(options=args, protocol=_VisualizerServer)
                pm.GetGlobalController().TriggerBreakRMIs()
    yield (idx, targetPartition)

// ─────────────────────────────────
results = rdd.mapPartitionsWithIndex(processPartition).collect()
for out in results:
    print(out)
```

Fig. 14. Spark-MPI parallel visualization using ParaViewWeb

The visualization processes use the MPI ranks to launch the ParaView server for remote Web-based visualization of the reconstructed dataset. We perform a conversion from the NumPy array to a VTK data structure before sending the dataset to the ParaView. Figure 15 shows the visualization of reconstructed TEM data in the ParaViewWeb client, Visualizer application, in a Web Browser.

To analyze the performance of our Spark-MPI approach, we have captured initial results on a small system with 128 GB of memory, and 16 Intel(R) Xeon(R) CPU E5-2640 v3 @ 2.60GHz processors for the 256x256x74 TEM demonstration that produced a 256x256x256 reconstructed dataset. We varied the number of Spark workers (from 1 to 12), which performs the reconstruction, and the number of MPI ranks (from 1 to 4), which implements the visualization. The total time of the ART algorithm went down to ~300 secs in our exploration, an improvement of 6x over the implementation in TomViz (see Fig. 16)

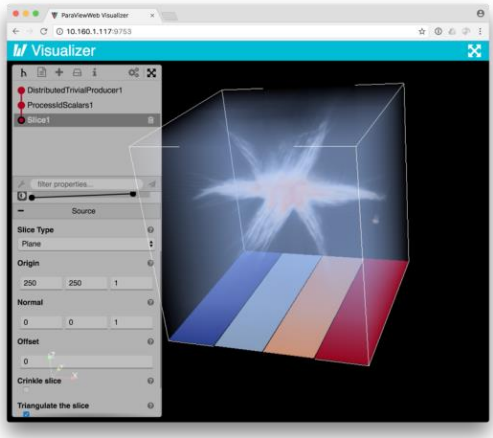

Fig. 15. Parallel visualization of reconstructed data in ParaViewWeb

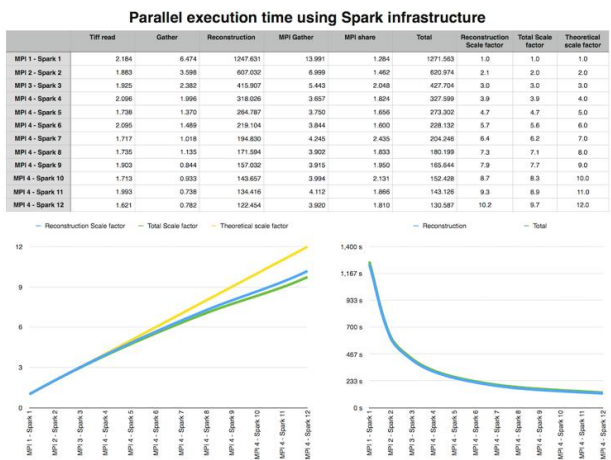

Fig. 16. Performance of reconstruction operation in Spark-MPI (time in secs).

## V. CONCLUSIONS

This paper presents the first of its kind hybrid Spark-MPI approach for building data-intensive high-performance data processing pipelines for experimental facilities. The approach is based on extending the Spark runtime environment with the Process Management Interface (PMI) server that complements the functionality of the Spark cluster manager for establishing the MPI-compliant inter-worker communication among distributed workers. The resulting integrated platform was demonstrated and accessed with two MPI applications, ptychographic and tomographic pipelines. The applications validated the conceptual approach and identified several directions. First, the Process Management Interface needs to be enhanced with the support of interactive MPI applications. This topic is generic and consistent with requirements of other MPI-based projects. Second, the Kafka Receiver can be augmented with interfaces to other data sources, such as ZeroMQ. Finally, the paper provides guidance for developing composite pipelines augmenting image reconstruction algorithms with deep learning analysis and feedback loop.

REFERENCES

[1] T. Hey, S. Tansley, K. Tolle, Eds. The Fourth Paradigm: Data-Intensive Scientific Discovery , Microsoft Research, 2009.

[2] K. K. Van Dam et al., "Chalenges in data intensive analysis at scientific experimental user facilities, " Handbook of Data Intensive Computing, Springer, 2011.

[3] P. Nugent et al., "Data and communications in basic energy sciences: creating a pathway for scientific discovery," DOE Workshop Report, 2012.

[4] Working Group. Accelerating Scientific Knowledge Discovery, DOE Workshop Report, 2013

[5] V. Sarkar et al., "Synergistic challenges in data-intensive science and exascale computing," DOE ASCAC Data Subcommittee Report, 2013

[6] J. Dean and S. Ghemawat. "MapReduce: simplified data processing on large clusters, " OSDI, 2004

[7] M. Zaharia et al., "Resilient distributed datasets: a fault-tolerant abstraction for in-memory cluster computing," NSDI, 2012.

[8] N. Marz and J. Warren, Big Data: Principles and Best Practices of Scalable Realtime Data Systems, Manning Publications, 2015.

[9] M. Zaharia et al., "Descretized streams: fault-tolerant streaming computation at scale," SOSP, 2013.

[10] MPI: A Message-Passing Interface Standard, Version 3.1, Message Passing Interface Forum, 2015

[11] Y. Low et al., "GraphLab: a new framework for parallel machine learning," UAI, 2010

[12] J. Dean et al., "Large scale distributed deep networks," NIPS, 2012

[13] M. Abadi et al., "TensorFlow: large-scale machine learning on heterogeneous distributed systems," Preliminary White Paper, 2015

[14] A. Nair et al., "Massively parallel methods for deep reinforcement learning," ICML, 2015

[15] G.Fox et al., "Towards an understanding of facets and examplars of Big Data applications,"  In Proc. of 20 Years of Beowulf, 2014

[16] G. Fox et al., "HPC-ABDS high performance computing enhanced Apache Big Data stack, " CCGrid

[17] C. Noel, J. Shi, and A. Feng, "Large scale distributed deep learning on hadoop clusters, " 2015 (unpublished)

[18] N. Malitsky, "Bringing the reconstruction algorithms to Big Data playforms, " NYSDS, 2016

[19] P. Balaji et al., "PMI: a scalable parallel process-management interface for extreme-scale systems,"  EuroMPI, 2010

[20] V. K. Vavilapalli et al., "Apache Hadoop YARN: yet another resource negotiator, " SoCC, 2013

[21] B. Hindman et al., "Mesos: a platform for fine-grained reource sharing in the data center, " NSDI, 2011

[22] M. Jette and M.Grondona, "SLURM: simple Linux utility for resource management, " UCRL-MA-147996 Rev 3, 2003

[23] J. Kreps, N. Narkhede, and J. Rao. "Kafka: a Distributed Messaging System for Log Processing,"  NetDB, 2011

[24] R. Hegerl and W. Hoppe, "Dynamische Theorie der Kristallstrukturanalyse durch Electronenbeugung im inhomogenen Primarstrahlwellenfeld," Phys. Chem. 7, 1970.

[25] J. M. Rodenburg et al., "Hard-X-Ray Lensless Imaging of Extended Objects, " Physical Review Letters 98, 2007

[26] P. Thibault et al., "High-Resolution Scanning X-ray Diffraction Microscopy, " Science 321, 2008.

[27] P. Thibault et al., "Probe retrieval in ptychographic coherent diffractive imaging, " Ultramicroscopy 109, 2009.

[28] K. Giewekemeyer, "A study on new approaches in coherent x-ray microscopy of biological speciments, " PhD thesis, Gottingen series in x-ray physics, Volume 5, 2011

[29] V. Elser, "Phase retrieval by iterated projections, " J. Opt. Soc. Am. A, 2003

[30] D. R. Luke, "Relaxed average alternating reflections for difraction imaging. Inverse Problems 21, 2005.

[31] S. Marchesini et al., "SHARP: A Distributed, GPU-Based Ptychographic Solver, " LBNL-1003977, 2016.

[32] B. Levin et al. "Data Descriptor: Nanomaterial Datasets to Advance Tomography in Scanning Transmittion Electron Microscopy, " Scientific Data, 2016.41

[33] TomViz, https://github.com/OpenChemistry/tomviz .